\documentclass[twocolumn,tighten]{aastex6}
\usepackage{hyperref,url}
\usepackage{natbib,graphicx,amsmath}

\shorttitle{Stellar variability and extended main sequences}
\shortauthors{Salinas et al.}

\newcommand{\dsct}{$\delta$~Sct\,}
\begin{document}

\title{The overlooked role of stellar variability in the extended main sequence of LMC intermediate-age clusters}

\author{R. Salinas\altaffilmark{1}$^{\star}$, M. A. Pajkos\altaffilmark{2}$^{\dagger}$, J. Strader\altaffilmark{3}, A. K. Vivas\altaffilmark{4} and R. Contreras Ramos\altaffilmark{5,6}}
          
\altaffiltext{1}{Gemini Observatory, Casilla 603, La Serena, Chile}
\altaffiltext{2}{Department of Physics and Astronomy, Butler University, Indianapolis, IN 46208, USA}
\altaffiltext{3}{Department of Physics and Astronomy, Michigan State University, East Lansing, MI 48824, USA}            
\altaffiltext{4}{Cerro Tololo Interamerican Observatory, National Optical Astronomy Observatory, Casilla 603, La Serena, Chile}
\altaffiltext{5}{Millennium Institute of Astrophysics, Av. Vicu\~na Mackenna 4860, 782-0436 Macul, Santiago, Chile}
\altaffiltext{6}{Instituto de Astrof{\'i}sica, Pontificia Universidad Cat\'olica de Chile, Av. Vicu\~na Mackenna 4860, 782-0436 Macul, Chile}
         \email{$^{\star}$ rsalinas@gemini.edu}

\begin{abstract}

Intermediate-age star clusters in the Large Magellanic Cloud show extended main sequence turn offs (MSTOs), which are not consistent with a canonical single stellar population. These broad turn offs have been interpreted as evidence for extended star formation and/or stellar rotation. Since most of these studies use single frames per filter to do the photometry, the presence of variable stars near the MSTO in these clusters has remained unnoticed and their impact totally ignored. We model the influence of Delta Scuti using synthetic CMDs, adding variable stars following different levels of incidence and amplitude distributions. We show that Delta Scuti observed at a single phase will produce a broadening of the MSTO without affecting other areas of a CMD like the upper MS or the red clump; furthermore, the amount of spread introduced correlates with cluster age as observed. This broadening is constrained to ages $\sim$ 1--3 Gyr when the MSTO area crosses the instability strip,  which is also consistent with observations. Variable stars cannot explain bifurcarted MSTOs or the extended MSTOs seen in some young clusters,  but they can make an important contribution to the extended MSTOs in intermediate-age clusters.

\end{abstract}

   
\section{Introduction}\label{sec:intro}

\let\thefootnote\relax \footnotetext{$^{\mathrm{\dagger}}$ CTIO/Gemini REU student.}

The existence of multiple stellar populations in Galactic globular clusters (GCs) is considered a widespread phenomenon \citep[][and references therein]{gratton12}, with only a few possible exceptions \citep[see][and references therein]{salinas15}. The presence of a second (and subsequent) populations is most likely associated to the ability of clusters to retain enriched material expelled to the ISM either from supernovae, AGB stars, fast-rotating stars or other types of interacting binaries \citep[e.g.][]{ventura01,bastian13}. Even though chemically distinct, the age difference that should exist between populations, between tens to a few hundred Myr depending on the origin of the enriched material,  is extremely difficult to measure given the ancient ($\sim$ 12 Gyr) nature of all Galactic GCs.

On the contrary, star clusters in the Large Magellanic Cloud (LMC) exhibit a wide range of ages from a few Myr up to clusters as old as the Milky Way GCs, with a distribution of masses overlapping the one in our Galaxy \citep[e.g.][]{baumgardt13}. The study of younger clusters may give us a glimpse of the conditions of early stellar evolution in the now old GCs. Specifically, in the case of young (tens to hundreds of Myrs) and intermediate-age (1--2 Gyr) clusters, in principle both the chemistry and age difference between populations can be measured giving access to a more complete knowledge of their star formation history.

The first clear evidence of multiple stellar populations in LMC clusters came from precise \textit{Hubble Space Telescope} (\textit{HST}) photometry of intermediate-age GCs revealing  extended main sequences \citep{mackey07,mackey08,milone09}, which was interpreted as extended star formation histories lasting a few hundreds of Myrs. \citet{milone09} argued that up to 75\% of their sample of 55 LMC clusters were not consistent with the single stellar population (SSP) hypothesis. The widespread nature of extended main sequences ruled out the possibility of them being the outcome of the mergers of two clusters with slightly different ages \citep[e.g.][]{goudfrooij09,milone09}. 

The straightforward interpretation of extended main sequences as signature of extended formation histories was quickly contended. An age spread visible at the MSTO level should also be visible at the red clump, but these do not show the expected spread in color and luminosity, being rather consistent with SSPs \citep[][but see \citealt{goudfrooij15} for a different view]{li14,bastian15}, also, in young massive clusters in the LMC, where the effect on the color-magnitude diagrams should be even clearer, the evidence for multiple populations remains controversial  \citep{bastian13b,niederhofer15,correnti15}.

Given the difficulties that the multiple stellar populations hypothesis face, stellar rotation has emerged as a more appealing explanation. Fast rotating stars with ages $\sim$ 1.5 Gyr, when viewed from different angles would produce the same extension of the MSTO, mimicking the effect of multiple stellar populations \citep{bastian09,yang13,brandt15}, although it has been claimed that rotation alone cannot fully reproduce the extended MSTO morphology \citep{girardi11}.

Here we study the contribution of yet another ingredient into the nature of extended MSs: the fact that the instability strip crosses the upper MS, MSTO and sub giant branch of intermediate-age clusters. Variable stars near the MSTO could produce an spurious luminosity and color spread if their pulsation properties, in particular amplitudes, are not properly taken into account. This is specially true when deriving CMDs from single or very few images per filter as is the usual case for \textit{HST} photometry \citep[e.g.][]{mackey07,milone09}. One attractive feature of this idea is that it would explain the absence of extended MSTOs in clusters older than $\sim$2.5 Gyr, given that the TO for these ages is redder than the instability strip, so no pulsations at the MSTO level would be produced. It will also explain that the color spreads are limited to the MSTO level and not affecting the red clump.

\subsection{Delta Scuti stars}\label{sec:dsct}

 \begin{figure}
   \centering
  \includegraphics[width=0.46\textwidth]{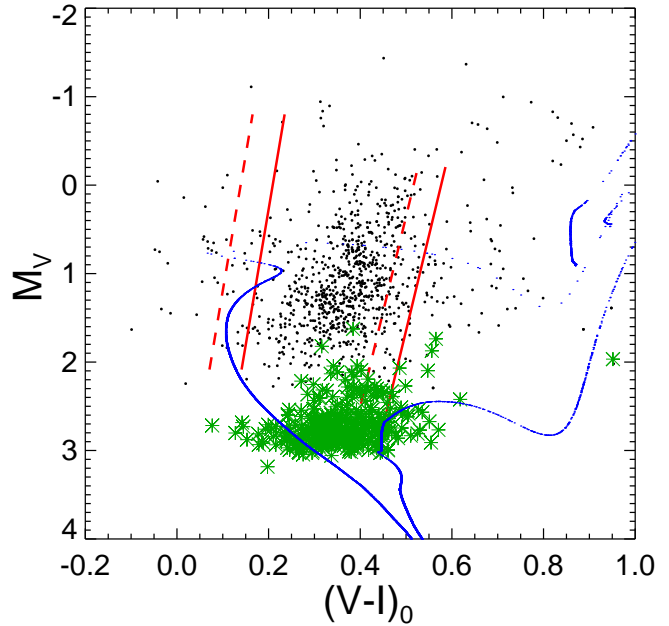}
\caption{The instability strip at the MSTO level. Dashed red lines indicate the empirical limits of the instability strip for Galactic \dsct from \citet{rodriguez01}. Solid lines indicate new limits broadly covering the colors of \dsct in the LMC from \citet{poleski10} (small black circles) and  Carina \citep[][green symbols]{vivas13}. Blue lines indicate BaSTI isochrones for 1 and 3 Gyrs with [Fe/H]=--0.5.}
         \label{fig:instability}
   \end{figure}

Delta Scuti (hereafter \dsct) are variable stars presenting both radial and non-radial pulsations produced by the well-known $\kappa$ mechanism \citep{baker62,cox80}, with periods between 30 minutes up to 8 hours. They are main sequence stars with spectral types from late A to early F and masses between 1.5 and 2.5 M$_{\odot}$ \citep[see e.g.][for a review]{breger00}.

Fig. \ref{fig:instability} shows two isochrones from BaSTI models \citep{pietrinferni04} of 1 and 3 Gyrs  with [Fe/H]=--0.5, representative of the intermediate-age cluster population in the LMC \citep[e.g.][]{grocholski06}. Overlaid, in dashed red lines, are the limits of the empirical Galactic lower instability strip at the MSTO level from  \citet{rodriguez01}. Since the \citet{rodriguez01} sample is mostly confined to solar neighborhood variables, their metallicity is close to solar. To more properly address the instability strip at lower metallicities we use photometry of \dsct in the body of the LMC from \citet{poleski10} (black symbols) and the Carina dSph \citep[][green symbols]{vivas13}. We define new empirical limits of the instability strip (solid red lines) encompassing the bulk  of the \dsct in these two galaxies, with the caveat that these samples are likely contaminated with \dsct of lower metallicities and probably metal-poor SX Phe variables, making its red limit uncertain.

Stars in the upper MS, MSTO and subgiant branch for stellar populations with $\sim$ 1 -- 3 Gyr will therefore experience pulsations. But unlike RR Lyrae stars, which lie at the intersection of the horizontal branch and the instability strip, not all MS stars inside the instability strip develop pulsations. The percentage of pulsating stars compared to the total number of stars inside the instability strip is known as the incidence. The incidence of $\delta$ Sct in the MW, from the study of the Solar neighborhood and open clusters is somewhere between a quarter and half of the stars within the instability strip \citep{breger00,poretti03,balona11}, although the majority of these have very small amplitudes of a few milimagnitudes. 

\begin{figure*}
    \includegraphics[width=\textwidth]{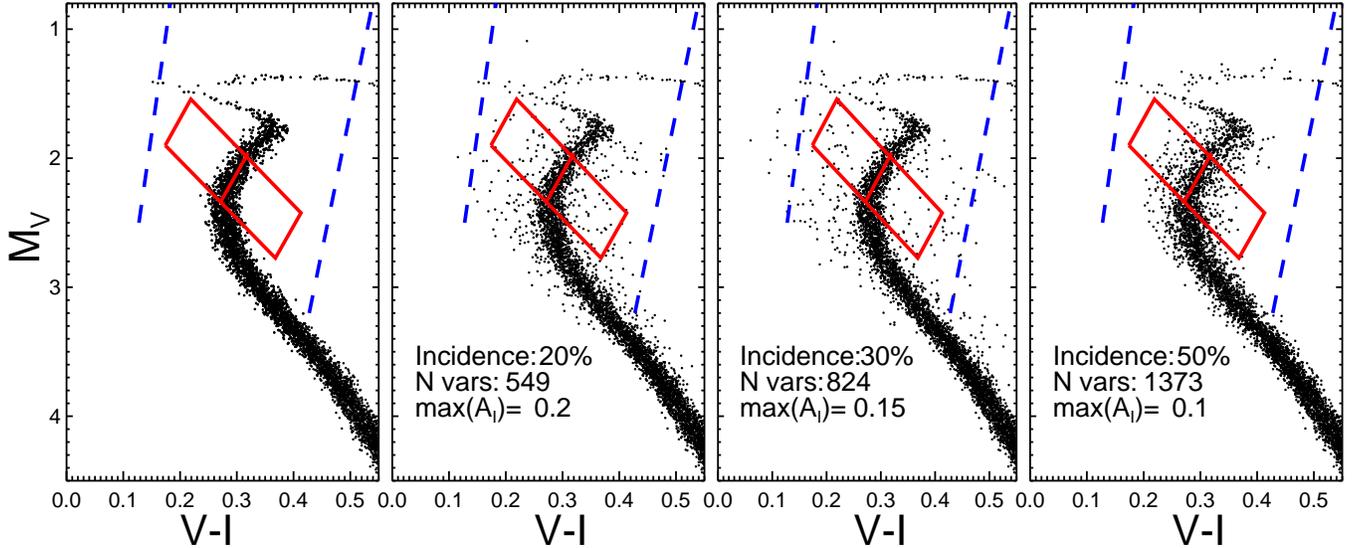}
\caption{Our models based on BaSTI synthetic CMDs. In all four panels a 1.5 Gyr old population with [Fe/H]=--0.5 was used. Blue dashed lines indicate the limits of the instability strip defined in Sect. \ref{sec:dsct}. Each panel indicates the \dsct incidence used, their maximum amplitude and the number of \dsct per cluster, while the leftmost panel shows the original synthetic CMD without variables. The red box indicates the section used to calculate the color spread introduced by the variables.}
\label{fig:sim_cmd}
\end{figure*}

Highly uncertain are the incidence values for \dsct in extragalactic systems because it is expected that a large part of them will not be detected given their low amplitudes, implying a lower incidence. In Carina, \citet{vivas13} report a lower limit of incidence of 8\%. However, the number of high amplitude \dsct in extragalactic systems like Carina and Fornax seems to be intrinsically higher than those observed in the field of the MW \citep{poretti08,vivas13}. In Carina, the peak of the amplitude distribution is $A_V$=0.5 mags \citep{vivas13} while in the LMC is $\sim$0.3 mags \citep{poleski10}. 

In this paper we explore the influence of \dsct on the morphology of the MSTO by using synthetic CMDs modified by the presence of these variable stars.

\section{The influence of Delta Scuti in a CMD}\label{sec:modellng}

\subsection{Incidence and amplitudes} \label{sec:incidence}
   
Since the great majority of extended MSTO have been detected using \textit{HST} imaging \citep[e.g.][]{milone09}, our models try to reproduce those CMDs. From DAOphot catalogs of \textit{HST}/ACS photometry of NGC 1846 (GO: 10595, PI: Goudfrooij) obtained from the Hubble Legacy Archive$^4$\footnote{$^4$\url{hla.stsci.edu}}, photometric errors at the MSTO level are of the order of 0.008 mag. Conservatively we assume an error of  0.01 mag per filter for the entire CMD.
Our models use as starting point BaSTI synthetic CMDs \citep{pietrinferni04} with [Fe/H]=--0.5, scaled to solar mixture and convective overshooting. Stars follow a luminosity function implied by the \citet{salpeter55} IMF, using a total number stars resulting in a population of $M_V\sim~-7.5$, typical for a fairly massive intermediate-age LMC cluster. Different ages between 0.75 and 2.5 Gyr were used (see below). 

The instability strip defined in Sect.~\ref{sec:dsct} was used to set the color boundaries from which stars would be selected to be modeled as variables. Stars within the instability strip were chosen randomly for values of the incidence between 10\% to 50\%. The selected stars were then modeled as variables using the cosine expansion of \citet{simon81},
\begin{equation}
I=A_0 + \sum_{i=1}^N A_i \cos(i\omega t + \phi_i),
\end{equation}
where $A_0$ is the ``static" magnitude coming from the CMD modeling described above.
    
Amplitudes and phases are usually grouped in the form $R_{ij}\equiv A_i/A_j$ and $\phi_{ij}\equiv\phi_i-i\phi_j$, and we took these values from the study of \dsct in the first OGLE bulge campaign of \citet{morgan98}. The \citet{morgan98} were preferred over the \citet{poleski10} values from OGLE-III since the former studied the Fourier parameters up to the 4th term, while the latter gives values for $R_{21}$ and $\phi_{21}$ only. In practice, for each star, values for the Fourier parameters were drawn from uniform distributions, $R_{21}=U(0.05,0.45)$, $R_{31}=U(0,0.2)$, $R_{41}=U(0,0.15)$ and, $\phi_1=U(0,0.15)$, $\phi_{21}=U(3.5,4.5)$, $\phi_{31}=U(1,3)$, $\phi_{41}=U(3.5,8)$, following the results of \citet{morgan98} and where we use the notation $U(x,y)$ to describe a uniform distribution with limits $x$ and $y$. The most relevant quantity is $A_1$ which dominates the total amplitude of the light variation. Its impact was studied drawing its distribution from several uniform distributions with maximum values for $A_1$ between 0.1 and 0.2. This is a conservative approach considering the largest amplitudes for \dsct in the LMC reach $\sim$ 0.8 mags in $I$ \citep{poleski10}.

  \begin{figure}
   \centering
  \includegraphics[width=0.46\textwidth]{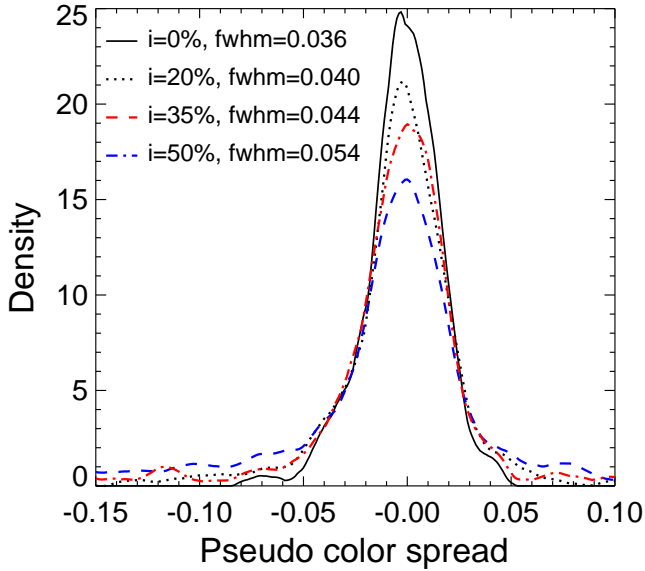}
      \caption{Distribution of pseudo-colors(ages) as function of incidence. In all cases a maximum amplitude $A_1=0.1$ was used. The sold black lines shows the spread when no variables are included.  }
         \label{fig:densities}
   \end{figure}
   
Since Fourier parameters exist only for the observations in $I$, light curves in $V$ were generated as scaled up versions of the $I$ light curves applying to $A_1$ the $A_V/A_I=1.7$ empirical factor given by \citet{rodriguez07}. The period distribution for these simulated variables was taken as a normal distribution centered in 0.075 days and with dispersion of 0.02 days following the results of \citet{poleski10} Finally, the modeled light curves were ``observed'' at one fixed time for each filter, that is,  taking for each variable a measurement from their light curve at a random phase and putting back these magnitudes into the CMDs.

The results from the modeling can be seen in Fig. \ref{fig:sim_cmd}, where three combinations of incidence and maximum amplitude are shown. The most salient result is that even though stars can develop pulsations everywhere within the instability strip, the color and magnitude shift of stars in the upper MS is mostly parallel to the MS incurring in little broadening of the MS width. On the contrary, \dsct near the TO experience color and luminosity changes almost perpendicular to the TO artificially broadening the MSTO when observed at a random phase.

In order to quantify the broadening of the MSTO area introduced by the \dsct, following \citet{goudfrooij11} we construct a parallelogram where one pair of the sides is roughly parallel to the isochrones, while the other pair is approximately perpendicular (see red boxes in Fig. \ref{fig:sim_cmd}). The parallelogram is located where the difference in the isochrones is the largest, therefore the minimum distance to the axis that is parallel to the isochrones can be considered as a ``pseudo-age", which in this case is introduced by the variable stars and does not represent a real age spread.  The distribution of these pseudo-color/ages as function of incidence with a fixed maximum amplitude $A_1=0.1$, can be seen in Fig. \ref{fig:densities}, where generalized histograms have been produced using an Epanechnikov kernel \citep{silverman86}. The models shows that the density of stars near its parent isochrone decreases with increasing incidence, and the distributions developed extended wings departing from the Gaussian errors. The FWHM of the distribution can increase up to a 50\% when an incidence of 50\% is used.

   \begin{figure}
   \centering
   \includegraphics[width=0.46\textwidth]{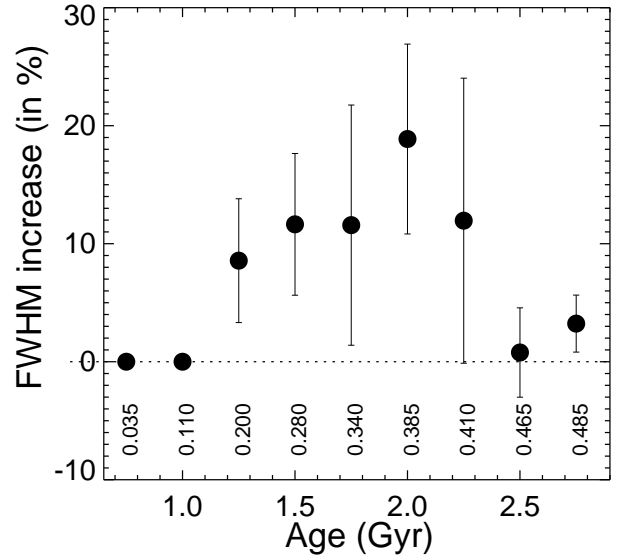}
   \caption{Distribution of MSTO spread as function of cluster age. All models have an incidence of 30\%. The increase and decrease of the FWHM with a peak at $\sim$2 Gyr broadly reproduces the observations of \citet{niederhofer16}. For each age the label indicates the $(V-I)_0$ color at the MSTO. }
         \label{fig:ages}
   \end{figure}
   
\subsection{MSTO spread as function of age}

As seen in Fig. \ref{fig:instability}, \dsct  will only affect the MSTO between ages $\sim$ 1 and 3 Gyr. For ages below 1 Gyr the TO point is too bright and \dsct will be only produced in the MS. For ages older than 3 Gyr the redder edge of instability strip is bluer than the MSTO so no pulsations will be developed in the MS (only in some possible blue stragglers).

We quantify the influence of \dsct as function of age running our model with fixed incidence of 30\% and maximum amplitude $A_1=0.2$, for ages from 0.75 to 2.75 Gyr with a step of 0.25 Gyr. The pseudo-age spread introduced at the MSTO is measured with the parallelogram approach introduced in the previous Section, centering it in each corresponding isochrone.

\citet{niederhofer16} used data from \citet{goudfrooij14} to argue that age spreads in the LMC clusters increase as function of cluster age, reaching a maximum at 1.5--1.7 Gyr, and then decreasing after  that age.

Figure \ref{fig:ages} shows the change in the FWHM of the pseudo age distributions as function of model age. These are the mean values and standard deviations from 50 models per age. In general it shows a behavior similar to the one found by \citet{niederhofer16}, although with a smaller amplitude. This behavior is easily explained as the MSTO region moves in and out of the instability strip making the highest number of stars to develop pulsations close to 2 Gyr. We stress, however, that absolute ages are model dependent, and different sets of isochrones will show differences of a couple hundred Myrs to define the edges of the instability strip.

\section{Summary and conclusions}

In this paper we explored the influence the \dsct pulsators have in the morphology of the MSTO in intermediate-age clusters in the LMC. Since the great majority of the photometry of these clusters comes from using merely 1 or 2 images per filter \citep[e.g.][]{mackey07,milone09,piatti14}, these variables have so far being undetected, and their role ignored.

We modeled an intermediate-age cluster using BaSTI isochrones, introducing \dsct populations of different amplitude distributions and incidence using \dsct shapes from \citet{morgan98}. Our model shows, 

\begin{itemize}
\item[--] a color spread is introduced by \dsct only at the MSTO level and  very little at the upper MS,
\item[--] this spread can be up to 50\% of the original value when a large number of variables are present,
\item[--]  the color spread is a function of cluster age resembling the pattern discovered by \citet{niederhofer16} , 
\item[--] no color spread will be introduced at the red clump, 
\item[--] no spread will be seen for cluster older than $\sim$ 3 Gyr.
\end{itemize}

There are several issues that variable stars \textit{cannot} explain, most notably the existence of bifurcated MSTOs in clusters like NGC 1846 \citep{mackey07} and the presence of extended MSTOs in young clusters like NGC 1856 \citep{milone15,correnti15}, where no influence of \dsct at the MSTO level is expected. Nevertheless, we have shown that \dsct will affect the MSTO morphology for clusters in the 1--3 Gyr age range, introducing a spread that can be misinterpreted as a departures from a simple stellar population, and therefore they cannot be ignored.

It is possible that the extended MSTO phenomenon in intermediate-age clusters is actually the result of a combination of an extended star formation, stellar rotation and stellar variability, but while the first two have been very difficult to disentangle \citep[e.g.][]{niederhofer16}, the impact of the latter, controlled by the incidence and the amplitude distribution of \dsct,  will be the easiest to properly assess once high spatial resolution time series images of these clusters become available.

\acknowledgements

We thank the referee for a helpful report which prompted a more realistic modeling. RS thanks Nate Bastian, Roger Cohen, Antonino Milone and Aldo Valcarce for useful conversations. This project was conducted in the framework of the CTIO REU Program, which is supported by the National Science Foundation under grant AST-1062976. JS acknowledges partial support from NSF grant AST-1514763 and the Packard Foundation. Based on observations made with the NASA/ESA Hubble Space Telescope, and obtained from the Hubble Legacy Archive, which is a collaboration between the Space Telescope Science Institute (STScI/NASA), the Space Telescope European Coordinating Facility (ST-ECF/ESA) and the Canadian Astronomy Data Centre (CADC/NRC/CSA). This work has made use of BaSTI web tools.

%\bibliography{letter}

\end{document}